\documentclass[letterpaper]{article}
\usepackage{spconf,amsmath,amssymb,graphicx}
\usepackage{booktabs,cite,enumitem,tabularx}
\usepackage[hidelinks]{hyperref}
\newcommand{\emailicon}{%
  \raisebox{0pt}[5bp][0pt]{\hbox{%
    \pdfliteral{q 0.35 w 0 0 7 5 re S 0 5 m 3.5 2 l 7 5 l S Q}%
    \kern7bp}}%
}
\newcommand{\IoU}{\operatorname{IoU}}
\title{TEMA: Evidence-Grounded Temporal Question Answering in Multi-Turn Multi-Audio Dialogs}
\name{Kaidi Yang$^{1,2}$\textsuperscript{\href{mailto:yangkaidi25@mails.ucas.ac.cn}{\emailicon}}, Hualei Wang$^{1,2}$,Zhaohui Wang$^{1,2}$,Chenxuan Wang$^{1,2}$,Hong Liu$^{1}$,Xiangdong Wang$^{1,\star}$}
\address{$^1$ Institute of Computing Technology, Chinese Academy of Sciences, Beijing, China.\\
$^2$ University of Chinese Academy of Sciences, Beijing, China.}
\newcommand{\footnoteONLYtext}[1]{%
  \begingroup
  \renewcommand{\thefootnote}{}%
  \footnotetext[0]{\small #1}%
  \endgroup
}
\begin{document}
\ninept
\maketitle

\begin{abstract}
Multi-turn, multi-audio temporal question answering requires models to track target events across follow-up questions, recording switches, and historical references, recovering complete instances and their boundaries for temporal calculation and comparison. We propose TEMA, which connects event perception with evidence-based answering through Route, specifying the audio scope, and Span, describing all relevant intervals as conditional audio captions. We construct TEMA-Dialog with 40,704 dialogs and per-turn evidence and answer supervision, and TEMA-Bench for joint evaluation of evidence and final answers. Training combines temporal grounding initialization, full-dialog supervised fine-tuning, and completeness-first Span-only GRPO. Experiments on Qwen2.5-Omni and AF-Next show improved temporal question answering, particularly event localization and cross-audio comparison. Reinforcement learning applied solely to evidence further improves interval recovery and answer accuracy.
\end{abstract}

\vspace{-0.35em}
\section{Introduction}
\label{sec:introduction}
\footnoteONLYtext{$^\star\,\,\text{Corresponding author. }$}
Large audio-language models (LALMs) unify sound recognition, audio captioning, and question answering, giving users a flexible way to interact with recordings. As queries focus on specific events, models must also understand when sounds occur, how long they last, and their temporal order. Answering when a particular sound occurs or how far apart two events are, for example, requires aligning linguistic descriptions with accurate event boundaries. Alignment between event semantics and temporal positions is therefore an important foundation of fine-grained audio understanding.

Prior work has strengthened localization through temporal representations and training data. Sridhar et al.~\cite{r1} construct temporal question-answering supervision from event timestamps. TimeAudio~\cite{r2} introduces time tokens and time-aware encoding, and constructs FTAR to support dense audio captioning, audio grounding, and timeline-based speech summarization. TimePro-RL~\cite{r3} and TEMPO~\cite{r4} further combine explicit temporal information with verifiable rewards to improve temporal alignment. TAG-Bench~\cite{dai2026tag} further evaluates the recovery of all query-matching intervals within a single recording, including queries with multiple occurrences. Their main training and evaluation settings involve individual recordings. Precise localization is usually posed as a single-turn query, with an emphasis on timestamped audio captioning and audio grounding.

\begin{figure}[!t]
\centering
\includegraphics[pagebox=cropbox,clip,width=\columnwidth]{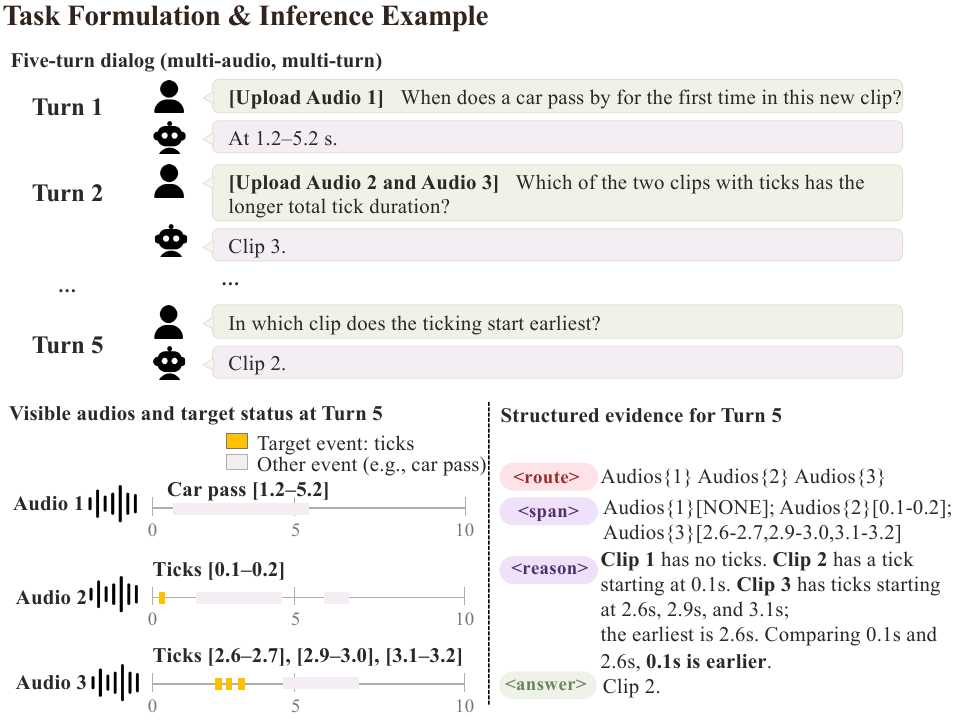}\par
\vspace{-1.0em}
\caption{Temporal QA and evidence in multi-audio dialog.}
\label{fig:task-example}
\vspace{-1.5em}
\end{figure}

In practical use, however, conversations often span multiple turns and may involve several recordings. As shown in Fig.~\ref{fig:task-example}, a user first asks for an event interval in one recording, then uploads additional recordings to compare the target sound's total duration and earliest occurrence. This interaction lets users refine questions based on previous analysis and requires models to maintain event--recording associations as queries change. AF-Chat~\cite{r5} provides dialog data supporting follow-up questions and references to earlier audio, with human experts evaluating factuality, helpfulness, and response depth. MUGEN~\cite{r6} examines acoustic-attribute identification and comparison across multiple recordings within a single turn. Although it includes temporal attributes such as duration and rhythm, it formulates questions as selection among candidate recordings and evaluates choice accuracy, without requiring event onset--offset intervals or explicit temporal calculations. These settings assess dialog understanding and cross-audio attribute judgments, but do not systematically test whether models can accurately recover event boundaries and correctly answer temporal questions across successive turns. For example, selecting the recording with the longer total ticking duration does not establish that a model has recovered every tick and its boundaries or calculated the total duration. A dataset and benchmark dedicated to such fine-grained temporal question answering in multi-turn, multi-audio dialogs remain lacking.

Multi-turn, multi-audio temporal question answering requires models to reason over the current question and dialog history, identify relevant recordings and events, recover instance boundaries on each timeline, and perform the requested calculations. Temporal localization must therefore adapt to targets and operations that change with context. Recent work has explored incorporating temporal information into reasoning. Echo~\cite{r7} generates intervals in its chain of thought and reintroduces the corresponding audio segments for re-listening. AF-Next's AF-Think-Time~\cite{r8} associates intermediate reasoning steps with timestamps to integrate evidence from long audio. These methods organize content through temporal references selected or generated during reasoning. However, introducing temporal references does not itself ensure that generated intervals accurately delimit the events required by the question. Multi-turn temporal questions also require evidence that accurately represents the current audio scope, event instances, and boundaries. Beyond temporal alignment rewards in TimePro-RL and TEMPO, we target complete query-conditioned evidence across dialog turns and recordings, jointly checking audio scope, instance counts, absence states, and boundaries. This is particularly important for counting and cumulative duration, where missing one occurrence can change the result. Event annotations should therefore define complete evidence for each question, enabling models to learn evidence acquisition before temporal calculation and comparison.

We study temporal question answering in multi-turn, multi-audio dialogs and propose TEMA, a framework based on query-conditioned temporal evidence. Recognizing the role of temporal evidence as an intermediate basis for reasoning, we construct a dataset and develop training methods that enable models to accurately acquire and use the information needed across follow-up questions, recording switches, and historical references. Our contributions are:
\begin{enumerate}[leftmargin=*,nosep,topsep=2pt]
\item \textbf{TEMA-Dialog.} We filter and process public audio data and event timestamps to construct a dataset with 40,704 dialogs and 198,195 response turns. To our knowledge, this is the first multi-turn, multi-audio temporal question-answering dataset. It covers event times, counts, and durations across follow-ups, cross-audio comparisons, and historical references, with complete temporal evidence, explanations, and answers per turn.
\item \textbf{Three-stage training for evidence acquisition and use.} Temporal grounding initialization aligns queries with event intervals; full-dialog SFT learns contextual evidence generation and answering; completeness-first Span-only GRPO directly optimizes instance completeness and boundary precision against event annotations. On TEMA-Bench, TEMA achieves higher overall QA than all compared baselines, including the open Qwen2.5-Omni-7B and AF-Next-Instruct and the proprietary Qwen3.5-Omni-Plus and Gemini 3.5 Flash.
\item \textbf{TEMA-Bench for joint evaluation of temporal evidence and final answers.} We retain model-generated history across 253 dialogs and 1,239 questions to assess evidence recovery and temporal question answering in natural multi-turn interaction, benchmarking precise temporal understanding in continuous dialog.
\end{enumerate}
We have released the code, dataset details and model weights.\footnote{\label{fn:tema-data}\small\url{https://github.com/KadeeYoung/TEMA}} 

\vspace{-0.35em}
\section{Methodology}
\label{sec:methodology}

\vspace{-0.5em}
\subsection{Task design and temporal evidence}
\label{sec:task}
\vspace{-0.35em}
Multi-turn, multi-audio temporal question answering uses currently visible recordings, dialog history, and the user question to answer questions involving event times, durations, counts, order, and temporal relations across recordings. Audio can be uploaded in two ways: \emph{upfront}, with all recordings provided at the first turn, or \emph{incrementally}, with new recordings introduced in later turns. Only uploaded audio is accessible at each turn.

The upper part of Fig.~\ref{fig:task-example} shows five turns with incremental uploads. As recordings and queries change, the model must identify target events and answer temporal questions. We organize audio questions into five families comprising 18 types, covering temporal attributes, multiple instances, and historical references.

\noindent\textbf{Task families.} \textbf{F1: Temporal localization and measurement} covers event localization, duration, and inter-event gaps. \textbf{F2: Event identification and verification} covers time-window identification, presence/absence, and false-premise correction. \textbf{F3: Within-audio temporal structure} covers instance counting and event ordering. \textbf{F4: Cross-audio retrieval and comparison} covers event retrieval, count comparison, earliest occurrence, exclusion retrieval, presence-and-count comparison, duration comparison, and per-audio counts. \textbf{F5: Multi-turn temporal reference} covers duration queries about recently mentioned or earlier-discussed event instances.

For consistent temporal supervision, one model sequentially generates Route, Span, Reason, and Answer at each turn. Route specifies the recordings requiring inspection or a judgment, including those where event absence must be verified. Span represents complete query-conditioned temporal evidence and can be viewed as structured conditional audio captioning. The question and history determine what to retrieve; Span records every matching instance and its onset--offset interval within Route's scope, marking recordings without matches as \texttt{[NONE]}. Reason selects instances, calculates, and compares; Answer gives the final response.

Turn 5 in Fig.~\ref{fig:task-example} illustrates the relation between evidence and operations. When comparing the earliest ticking occurrence, Route covers three recordings. Span records the absence of ticking in audio 1 and all ticking intervals in audios 2 and 3, after which Reason selects the earliest onset, 0.1\,s. Although this question needs only the earliest instance, Span retains all instances to support different temporal operations.

\vspace{-0.5em}
\subsection{Constructing the TEMA-Dialog dataset}
\label{sec:data}
\vspace{-0.35em}
\begin{figure}[!t]
\centering
\includegraphics[pagebox=cropbox,clip,width=\columnwidth]{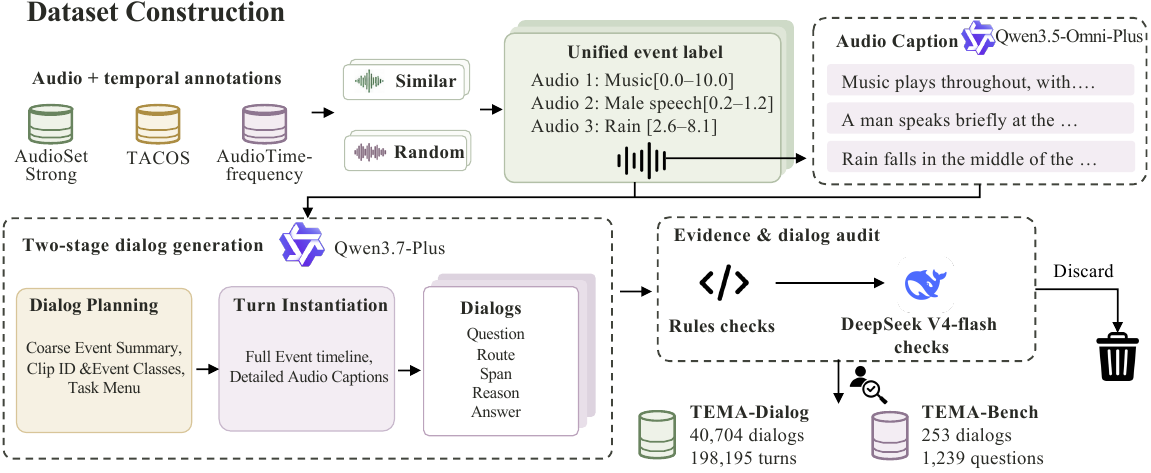}\par
\vspace{-1.0em}
\caption{Construction pipeline for TEMA-Dialog and TEMA-Bench.}
\label{fig:data-construction}
\vspace{-1.5em}
\end{figure}
Per-turn evidence requires accurate event annotations and consistent dialog content. We group recordings by event categories and intervals, then generate connected questions and answers (Fig.~\ref{fig:data-construction}).

\noindent\textbf{Preparing event annotations.} We integrate AudioSet Strong~\cite{r9}, TACOS~\cite{r10}, and AudioTime-frequency~\cite{r11} into unified event tables. Existing onset--offset intervals are retained. We map TACOS segment descriptions to AudioSet categories and use its ontology to normalize label granularity. AudioTime-frequency uses AudioSet-style event names, overlays events on silence, and provides counts and onsets. We complete intervals by estimating offsets from waveform amplitude and consecutive silence.

Beyond temporal localization, counting requires verified correspondence between event counts and individual intervals. From the 456 AudioSet Strong categories used in this project, we manually select 202 countable candidates and uniformly sample relevant recordings. Qwen3.5-Omni-Plus predicts counts and timestamps. We select core events using agreement between predicted and annotated counts and interval intersection-over-union (IoU), then expand to recordings containing these events and repeat verification. Counting questions use only verified audio--event records. AudioTime-frequency supplies supervision through its existing counts and onsets and the completed offsets.

\noindent\textbf{Audio grouping and dialog generation.} For cross-audio retrieval and temporal comparison, we use two grouping methods (Fig.~\ref{fig:data-construction}). Similar groups retrieve candidates using Jaccard similarity over event labels and filter them by minimum pairwise CLAP~\cite{r12} similarity within each group, supporting instance discrimination among similar sounds. Random groups introduce differences across acoustic scenes to diversify comparison relations. Qwen3.5-Omni-Plus generates captions for question wording.

To reduce factual and temporal errors in one-pass generation, Qwen3.7-Plus first plans dialogs, then instantiates each turn. Planning uses coarse event information, audio IDs, and a valid question-type--target-event--audio menu to arrange questions, upload order, and historical references; instantiation combines full event tables and detailed audio captions to generate each question and its four-part response. Event tables constrain instances, intervals, and operation results, while captions only guide wording, preserving consistency with temporal annotations across turns.

\noindent\textbf{Dialog auditing and dataset scale.} Generated dialogs undergo rule-based and model-based auditing. Rules check Route and Span against event tables; DeepSeek V4-flash further audits factual wording, audio IDs, and references. Samples that fail are discarded. We also manually review 100 sampled dialogs, with conclusions agreeing with the model-based audit.

TEMA-Dialog and TEMA-Bench draw recordings from the source datasets’ training and test splits, respectively. TEMA-Dialog comprises 40,704 dialogs, 198,195 response turns, and 75,561 recordings, with 1–6 recordings and 2–13 turns per dialog. Multi-audio dialogs account for 78.80\%; 23,858 dialogs use incremental uploads and 16,846 use upfront uploads. TEMA-Bench Test253 comprises 253 dialogs and 1,239 questions, including 209 multi-audio dialogs, with additional manual review and correction. See Footnote~\ref{fn:tema-data} for dataset details.

\begin{table*}[t]
\centering
\caption{Natural multi-turn results and ablations on Test253; gold-evidence diagnostics below. I/S/G: temporal grounding initialization, dialog SFT, and Span-only GRPO. All scores are percentages; family columns report QA. Evidence metrics follow Sec.~\ref{sec:setup}. ---: not evaluated. Bold: best available natural score. Percentage-point changes use displayed rounded scores.}
\label{tab:results}
\small
\setlength{\tabcolsep}{2.5pt}
\begin{tabular*}{\textwidth}{@{\extracolsep{\fill}}llrrrrrrrrr@{}}
\toprule
Model & Training & \multicolumn{2}{c}{Overall} & \multicolumn{2}{c}{Evidence} & \multicolumn{5}{c}{Task-family QA} \\
\cmidrule(lr){3-4}\cmidrule(lr){5-6}\cmidrule(lr){7-11}
& & QA & QA+T & Span F1 & H & F1  & F2  & F3  & F4  & F5  \\
\midrule
Qwen2.5-Omni-7B & Original & 46.09 & 38.90 & --- & --- & 26.05 & 76.89 & 54.84 & 22.67 & 33.33 \\
AF-Next-Instruct & Original & 51.98 & 46.57 & --- & --- & 39.07 & 72.79 & 80.65 & 29.78 & 36.11 \\
TEMPO multitask-RL & Original & 41.89 & 19.77 & --- & --- & 36.87 & 62.42 & 35.48 & 12.89 & 33.33 \\
Qwen3.5-Omni-Plus & Original & 57.71 & 46.49 & --- & --- & 28.04 & \textbf{89.42} & \textbf{87.10} & 52.44 & 5.56 \\
Gemini 3.5 Flash & Original & 49.72 & 39.39 & --- & --- & 21.41 & 80.78 & 53.23 & 48.00 & 11.11 \\
\midrule
TEMA (AF-Next) & I+S & 58.03 & 47.22 & 25.23 & 33.33 & 39.96 & 78.19 & 72.58 & 51.56 & 41.67 \\
TEMA (AF-Next) & I+S+G & 58.84 & 49.88 & 39.52 & 33.66 & 40.18 & 78.83 & 75.81 & \textbf{53.78} & 38.89 \\
TEMA (Qwen2.5-Omni) & S & 57.14 & 48.99 & 32.16 & 27.36 & 53.20 & 71.92 & 82.26 & 28.44 & 52.78 \\
TEMA (Qwen2.5-Omni) & I+S & 68.93 & 58.27 & 52.30 & 42.78 & 61.37 & 82.51 & 80.65 & 53.33 & 66.67 \\
TEMA (Qwen2.5-Omni) & I+S+G & \textbf{69.81} & \textbf{58.76} & \textbf{58.65} & \textbf{44.39} & \textbf{62.03} & 84.23 & \textbf{87.10} & 51.11 & \textbf{69.44} \\
\midrule
\multicolumn{11}{l}{\textit{Supplied gold Route/Span: diagnostic only}} \\
Qwen2.5-Omni-7B & Original & 82.81 & 81.52 & --- & --- & 89.62 & 87.69 & 72.58 & 62.22 & 80.56 \\
TEMA (Qwen2.5-Omni) & I+S & 96.45 & 95.80 & --- & --- & 95.36 & 99.35 & 98.39 & 92.44 & 94.44 \\
\bottomrule
\end{tabular*}\par
\vspace{-1.1em}
\end{table*}

\vspace{-0.5em}
\subsection{Three-stage training}
\label{sec:training}
\vspace{-0.35em}
Building on TEMA-Dialog, we develop a training framework for evidence-grounded temporal question answering. The baseline trained with dialog supervision alone still shows limited fine-grained temporal perception, motivating dedicated training for temporal evidence. We adopt three stages: first, learning query--interval alignment using the AudioGrounding subset of TimeAudio's FTAR; then, full-dialog SFT on TEMA-Dialog to learn contextual evidence generation and use; and finally, Span-only GRPO to improve instance completeness and boundary precision.

\noindent\textbf{Temporal grounding initialization (I).} We use 98,401 positive examples from the AudioGrounding subset of TimeAudio's FTAR~\cite{r2}. Given one recording and a query, token cross-entropy supervises a Span response containing all relevant intervals, establishing query-to-interval alignment for dialog evidence generation.

\noindent\textbf{Full-dialog supervised fine-tuning (S).} TEMA-Dialog supervises the complete Route, Span, Reason, and Answer at each turn with reference dialog history. The model learns to infer audio scope from the question and history, generate evidence, select instances, and perform temporal operations.

\noindent\textbf{Completeness-first Span-only GRPO (G).} Reinforcement learning for audio question answering requires reliable verification. R1-AQA~\cite{r17} and Audio-Thinker~\cite{r18} verify selection-based answers by choice matching. TEMA's open-ended answers require resolving events, instance references, and recordings: equivalent facts may differ in wording, while identical values may refer to different instances. String or value matching is therefore insufficient. An LLM judge adds cost without ensuring reward reliability or consistency.

TEMA's temporal evidence provides a more direct verification target. Route and Span enable direct verification of scope, instance counts, absence states, and endpoints against event tables. Given the evidence-to-answer mapping learned through dialog SFT and the reference-evidence diagnostics (Sec.~\ref{sec:analysis}), we hypothesize that improving evidence accuracy and completeness can improve answering.

Span-only GRPO~\cite{r15} generates and optimizes only Route and Span, stopping at \texttt{</span>}, without generating or directly rewarding Reason or Answer. We jointly evaluate event sets and boundaries because missing or extra instances affect counts, cumulative durations, and ordinal selections.

For generated evidence $y$, $H(y)=1$ iff the predicted audio scope, per-recording instance counts, and \texttt{[NONE]} states match the reference, with all onsets and offsets within 0.1\,s under within-audio one-to-one matching; otherwise, $H(y)=0$. A continuous score $D(y)\in[0,1]$ distinguishes candidates that are not yet complete. The reward is
\begingroup
\setlength{\abovedisplayskip}{5pt plus 1pt minus 1pt}
\setlength{\belowdisplayskip}{5pt plus 1pt minus 1pt}
\setlength{\abovedisplayshortskip}{3pt plus 1pt minus 1pt}
\setlength{\belowdisplayshortskip}{4pt plus 1pt minus 1pt}
\begin{equation}
r(y)=\begin{cases}
0.9H(y)+0.1D(y),&y\text{ is valid evidence},\\
-0.1,&\text{otherwise}.
\end{cases}
\label{eq:reward}
\end{equation}
\endgroup
Complete candidates score at least 0.9 and valid incomplete candidates at most 0.1, prioritizing completeness over partial correctness.

The continuous score combines audio scope, instance matching, and event absence. For a predicted interval $p$ and reference interval $g$, matching quality is
\begingroup
\setlength{\abovedisplayskip}{5pt plus 1pt minus 1pt}
\setlength{\belowdisplayskip}{5pt plus 1pt minus 1pt}
\setlength{\abovedisplayshortskip}{3pt plus 1pt minus 1pt}
\setlength{\belowdisplayshortskip}{4pt plus 1pt minus 1pt}
\begin{equation}
\begin{aligned}
q(p,g)&=0.7\IoU(p,g)\\
&\quad+0.3\exp\!\left(-\frac{|p_s-g_s|+|p_e-g_e|}{2\tau(g)}\right),\\
\tau(g)&=\operatorname{clip}\!\left(0.2(g_e-g_s),0.1,0.5\right),
\end{aligned}
\label{eq:quality}
\end{equation}
\endgroup
where $s,e$ denote onset and offset in seconds, and $\operatorname{clip}$ bounds the error scale to 0.1--0.5\,s. We perform maximum-weight Hungarian matching independently within each recording shared by reference and prediction. With matched quality sum $Q$ and reference/predicted nonempty interval counts $n,\widehat n$, the soft instance score is $F_{\mathrm{occ}}=2Q/(n+\widehat n)$ for $n>0$.

Let $R^*,\widehat R$ denote the reference and predicted audio sets. Set F1 measures their scope agreement: $M=2|R^*\cap\widehat R|/(|R^*|+|\widehat R|)$. Let $Z$ contain reference recordings without matching events. The absence score $N_{\mathrm{acc}}$ is the fraction of $Z$ correctly included in the predicted scope and explicitly marked \texttt{[NONE]}; it equals one when $Z$ is empty. Finally,
\begingroup
\setlength{\abovedisplayskip}{5pt plus 1pt minus 1pt}
\setlength{\belowdisplayskip}{5pt plus 1pt minus 1pt}
\setlength{\abovedisplayshortskip}{3pt plus 1pt minus 1pt}
\setlength{\belowdisplayshortskip}{4pt plus 1pt minus 1pt}
\begin{equation}
D=M\begin{cases}
0.8F_{\mathrm{occ}}+0.2N_{\mathrm{acc}},&n>0,\ |Z|>0,\\
F_{\mathrm{occ}},&n>0,\ |Z|=0,\\
N_{\mathrm{acc}},&n=0.
\end{cases}
\label{eq:continuous}
\end{equation}
\endgroup
\vspace{-0.75em}
\section{Experiments}
\label{sec:experiments}
\vspace{-0.5em}
\subsection{Experimental setup}
\label{sec:setup}
\vspace{-0.35em}
\noindent\textbf{Implementation details.} We evaluate the training framework on Qwen2.5-Omni-7B~\cite{r13} and AF-Next~\cite{r8}, both using the same temporal grounding data and TEMA-Dialog. The first two stages train LoRA~\cite{r14} adapters in the language model and audio-projection layers with rank 64, alpha 128, and dropout 0.05, keeping the base weights and audio encoder frozen.

Each backbone's SFT model screens its own GRPO questions. Taking Qwen2.5-Omni as an example, we sample eight evidence candidates per question from 7,292 questions, retaining 1,034 with both a completely correct and a valid but incomplete candidate. Each step processes four questions with four resampled candidates each. GRPO trains new language-model LoRA adapters with rank 16 and alpha 32, keeping the audio encoder and projector frozen.

\noindent\textbf{Evaluation protocol.} We evaluate all five families on Test253 using upload-order recordings and model-generated history, with greedy decoding capped at 1,024 tokens per turn for all models. Comparators include the open models Qwen2.5-Omni-7B, AF-Next-Instruct, and TEMPO multitask-RL, and the proprietary Qwen3.5-Omni-Plus and Gemini 3.5 Flash. Baselines in Table~\ref{tab:results} answer directly; TEMA and its ablations generate Route, Span, Reason, and Answer. We score all final answers and evidence only for structured outputs.

\noindent\textbf{Metrics.} Our questions concern verifiable event properties, including identity, occurrence counts, temporal order, and timing. Given the question, reference history, and event annotations, DeepSeek V4-flash judges non-temporal correctness, including event identity, counts, order, and recording/instance selection, and extracts required and optional temporal claims. Deterministic rules verify temporal values with an inclusive 0.1\,s tolerance. Within-event time points and explicitly stated subintervals are checked for containment. QA measures whether the final Answer satisfies the question’s requirements; QA+T additionally checks all volunteered temporal claims. Reference histories are used only for scoring, and scores are aggregated across turns. Evidence metrics are Span micro F1@0.5, pooling TP/FP/FN from within-audio one-to-one matching at IoU $\ge0.5$, and H@0.1 for complete evidence (Sec.~\ref{sec:training}).

\vspace{-0.5em}
\subsection{Main results}
\label{sec:results}
\vspace{-0.35em}
TEMA achieves 69.81\%/58.84\% QA on Qwen2.5-Omni/AF-Next, also outperforming non-TEMA models in QA+T (Table~\ref{tab:results}). Relative to original Qwen2.5-Omni, localization and measurement (F1) increases from 26.05\% to 62.03\%, cross-audio comparison (F4) from 22.67\% to 51.11\%, and historical reference (F5) also improves substantially. Qwen3.5-Omni-Plus leads in event identification and verification (F2) but remains well below TEMA in F1, highlighting the value of learning event intervals and their use for precise temporal understanding.

\vspace{-0.5em}
\subsection{Training-stage ablations}
\label{sec:ablations}
\vspace{-0.35em}
\textbf{Temporal grounding initialization.} Under identical dialog supervision, I+S improves Qwen2.5-Omni's QA and Span F1 by 11.79 and 20.14 percentage points, respectively, while also improving cross-audio comparison. This supports learning query-to-interval alignment before contextual evidence use.

\noindent\textbf{Span-only GRPO.} Adding GRPO improves Span F1 by 6.35/14.29 percentage points and QA by 0.88/0.81 points on Qwen2.5-Omni/AF-Next. Without directly rewarding Reason or Answer, these gains support our hypothesis (Sec.~\ref{sec:training}) that optimizing verifiable evidence after dialog supervision can improve answering.

\vspace{-0.5em}
\subsection{Temporal-evidence analysis}
\label{sec:analysis}
\vspace{-0.35em}
\textbf{The value of accurate evidence for answering.} To isolate evidence use, we supply reference Route and Span per turn, retaining model-generated history and excluding reference Reason and Answer. Qwen2.5-Omni's I+S model gains 27.52 percentage points in QA, indicating that it can use accurate temporal facts effectively. Even without explicit explanations of Route and Span, the original model reaches 82.81\%. These results support evidence-based answering while revealing room to improve evidence acquisition.

\noindent\textbf{Interval improvement and complete answering.} Evidence reinforcement learning improves interval recovery and overall answering on both backbones. However, Qwen2.5-Omni's cross-audio comparison drops from 53.33\% with I+S to 51.11\% with I+S+G, indicating that interval gains do not consistently translate into better answers across temporal operations. Future work should use turn-level error analysis to improve instance coverage, boundary precision, and cross-turn evidence consistency.

\vspace{-0.45em}
\section{Conclusion}
TEMA connects event perception with multi-turn, multi-audio answering through complete query-conditioned temporal evidence, supported by TEMA-Dialog for per-turn supervision and TEMA-Bench for joint evaluation. Three-stage training improves temporal QA on both backbones, with reinforcement learning on verifiable evidence alone further improving interval recovery and answering, supporting precise temporal understanding across multi-turn, multi-audio dialogs.

\clearpage
\bibliographystyle{IEEEbib}
\bibliography{references}
\end{document}